\RequirePackage{fix-cm}
\documentclass[twocolumn]{svjour3}          
\smartqed  
\usepackage{graphicx}
\usepackage{amsmath}
\usepackage{amssymb}
\usepackage{color}
\usepackage{braket}
\begin{document}

\title{A versatile design for resonant guided-wave parametric down-conversion sources for quantum repeaters}


\author{Benjamin Brecht \and
        Kai-Hong Luo \and
		Harald Herrmann \and
		Christine Silberhorn
}


\institute{Benjamin Brecht \and Kai-Hong Luo \and Harald Herrmann \and
		   Christine Silberhorn \at
           Integrated Quantum Optics, University of Paderborn, Warburger Strasse 100, 33098 Paderborn, Germany
            \emph{Present address: Clarendon Laboratory, Department of Physics, University of Oxford, Parks Road, OX1 3PU, United Kingdom} of Benjamin Brecht  
		  \email{benjamin.brecht@physics.ox.ac.uk}
}

\date{Received: date / Accepted: date}

\maketitle

\begin{abstract}
Quantum repeaters - fundamental building blocks for long-distance quantum communication - are based on the interaction between photons and quantum memories. The photons must fulfil stringent requirements on central frequency, spectral bandwidth and purity in order for this interaction to be efficient. We present a design scheme for monolithically integrated resonant photon-pair sources based on parametric down-conversion in nonlinear waveguides, which facilitate the generation of such photons. We investigate the impact of different design parameters on the performance of our source. The generated photon spectral bandwidths can be varied between several tens of MHz up to around $1\,$GHz, facilitating an efficient coupling to different memories. The central frequency of the generated photons can be coarsely tuned by adjusting the pump frequency, poling period and sample temperature and we identify stability requirements on the pump laser and sample temperature that can be readily fulfilled with off-the-shelve components. We find that our source is capable of generating high-purity photons over a wide range of photon bandwidths. Finally, the PDC emission can be frequency fine-tuned over several GHz by simultaneously adjusting the sample temperature and pump frequency. We conclude our study with demonstrating the adaptability of our source to different quantum memories. 
\keywords{Quantum optics \and Parametric down-conversion \and Quantum repeater}
\end{abstract}
\section{Introduction}
\label{intro}
Quantum communication is a means to exploit quantum mechanical principles to realise provably secure communication between distant parties. However, in any practical realisation, transmission losses in optical fibres limit the secure distance of quantum communication applications. This limitation can be overcome by utilising so-called quantum repeaters \cite{Briegel:1998wp}, which facilitate long-distance entanglement distribution and thus enable long-distance secure quantum communication (for a review see \cite{Sangouard:2011bp}). Quantum repeaters are based on the interaction between photons and quantum memories. Thus, they require the generation of narrowband photons that are compatible with an atomic transition in the quantum memory, ideally paired with partner photons that are suited for transmission via telecommunication fibres. The transmitted photon is used to establish entanglement between separate -- potentially dissimilar -- nodes of the quantum repeater, e.g. via the approach outlined in \cite{Simon:2007dc}. Note that, in this case, the generated photon pairs do not have to be in an entangled state themselves, since the two repeater nodes are entangled conditioned on the detection of a single photon behind a beamsplitter. Depending on the physical realisation of the quantum memory, the relevant atomic transitions are in the visible to near-infrared wavelength regime and their bandwidths typically range from few MHz to several GHz (see \cite{Simon:2010kl,Bussieres:2013br} and citations therein). 

In this paper, we present a design tool kit for photon-pair sources based on monolithically integrated, doubly resonant parametric down-conversion (PDC) in periodically poled lithium niobate (PPLN) waveguides. The strength of our robust design, as we will show, is that it can be adapted to a wide range of quantum memories. Thus, a single material and design platform is sufficient to realise large-scale quantum repeaters with dissimilar nodes, which lends itself to the rapid commercialisation of the field of quantum communication and the associated need for standardised fabrication procedures.

Typically, photons generated in a PDC source feature spectral bandwidths on the order of hundreds of GHz and more, due to the broad phasematching of the process. Although it is possible to deploy spectral bandpass filtering to narrow down the photons, the greater part of the generated PDC signal is lost during this process. This has two implications. First, the pump power is not efficiently used, which can cause prohibitive power consumption of large-scale applications with many sources. Second, the total generation rate of photon pairs cannot be made arbitrarily high, since higher-order photon terms start to pollute the photon-pair characteristics for generation probabilities in excess of few percents. This implies that external filtering of broadband PDC will generally result in low overall photon generation rates. A more appealing way towards generating narrow band photons is to directly manipulate the PDC emission spectrum by resonance enhancement within a cavity, also referred to as optical parametric oscillator below threshold. Typical resonant PDC sources reported so far are based on PDC in bulk crystals, which are placed inside a cavity \cite{Ou:1999wg,Kuklewicz:2006gq,Bao:2008hh,Scholz:2009bd,Hockel:2011hh,Wang:2008gf,Nielsen:2009fa,Wolfgramm:2011gf,JeronimoMoreno:2010kq,Chuu:2012ik,Fekete:2013kr,Zhou:2014fj}. There, bandwidths of $2\,$MHz have been achieved \cite{Fekete:2013kr}. However, bulk-crystal sources have disadvantages that limit their usefulness in large-scale quantum networks: first, they typically suffer from low pair production rates; second, they are not seamlessly integratable into a fibre-network architecture; third, most current realisations require sensitive cavity-locking schemes. 

An alternative kind of sources, which potentially overcomes these problems, is based on monolithically integrated resonant structures. One recent example of such a source is a whispering gallery mode resonator made from magnesium-doped lithium niobate, which features a narrow linewidth, a large tunability and single-longitudinal-mode emission \cite{Fortsch:2013ix}. The downside of this source is that it is not integrated. Another approach deployed a PPLN waveguide with dielectric mirror coatings on the waveguide end facets, which generated nearly degenerate photon pairs at telecommunication wavelengths \cite{Pomarico:2009dy}. Although integrated, this source did not feature single-longitudinal-mode emission due to the degeneracy of the PDC pair photons. To overcome this problem, frequency and polarisation non-degenerate PDC can be exploited \cite{PomaricoE:2012ew}.

Recently, we demonstrated a narrowband photon-pair source based on this scheme in a PPLN waveguide, which featured a photon spectral bandwidth of $60\,$MHz and generated photons at $890\,$nm and $1320\,$nm, designed to be used in conjunction with a neodymium-based quantum memory \cite{Luo:2015vz}. In addition, the source exhibited a very high spectral brightness of $3\times10^4$ pairs/(s$\,$mW$\,$MHz), making it an energy-efficient source for largescale networks. Based on these results, we demonstrate how this versatile and very general source design can be adapted to be compatible with different quantum memories. We will identify the influence of different fabrication parameters on the source performance and discuss the potential universality of our design for large-scale hybrid quantum networks comprising dissimilar building blocks. 

\section{Operational principle of our source}
\label{sec:1}
\begin{figure}
	\centering
	\includegraphics[width=.7\linewidth]{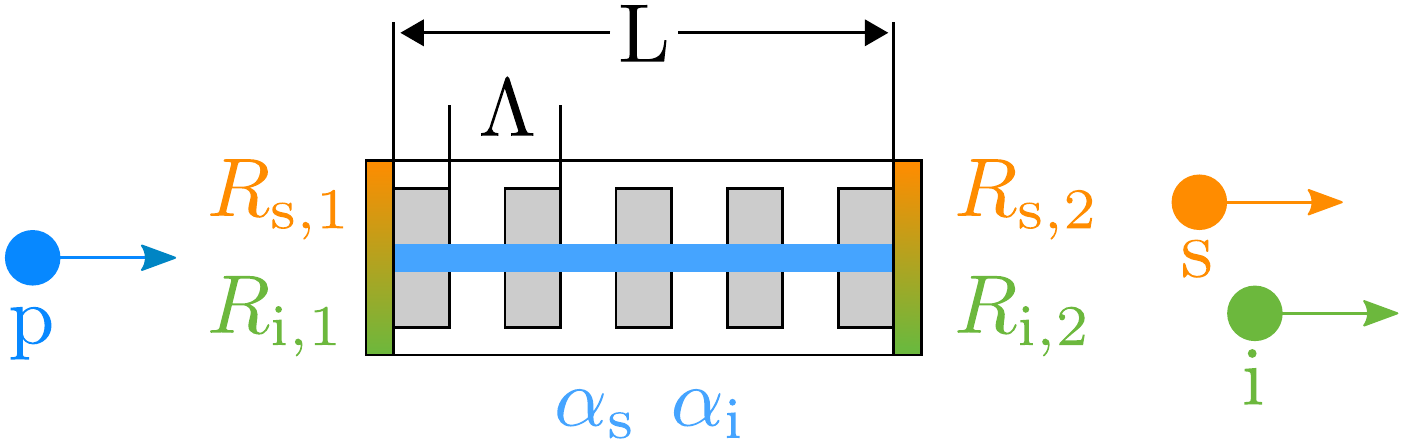}
	\caption{Our source design comprises a periodically poled waveguide of length $L$, whose endfacets are coated with highly reflective dielectric coatings. This structure forms a doubly-resonant cavity for the signal and idler photons that are generated in the process of PDC, when a pump photon decays inside the waveguide into a photon pair.}
	\label{fig:source_layout}
\end{figure}

Our source design is sketched in Fig.~\ref{fig:source_layout}. A PPLN waveguide of length $L$ with a poling period of $\Lambda$, which facilitates quasi-phasematching of the desired frequencies is at the heart of the design. In the PDC process, an ordinarily-polarised photon from a bright pump field decays into a pair of cross-polarised photons, customary labeled signal (ordinary polarisation) and idler (extra-ordinary polarisation), inside the waveguide. The process conserves energy and momentum, such that
\begin{equation}
	\omega_\mathrm{p} = \omega_\mathrm{s} + \omega_\mathrm{i}
	\label{eq:energy}
\end{equation} 
and 
\begin{equation}
	\beta_\mathrm{p} = \beta_\mathrm{s} + \beta_\mathrm{i} +
	\frac{2\pi}{\Lambda},
	\label{eq:phasematching}
\end{equation}
where the $\omega_j$ with $j={\mathrm{p},\mathrm{s},\mathrm{i}}$ denote the angular frequencies of the involved fields, and the $\beta_j=n_j(\omega_j)\omega_j/c$ are the respective propagation constants, with $n_j(\omega_j)$ being the effective refractive index of the waveguide mode and $c$ the speed of light. Dielectric coatings with reflectivities $R_\mathrm{s,1}$ and $R_\mathrm{s,2}$ for the signal and $R_\mathrm{i,1}$ and $R_\mathrm{i,2}$ for the idler, respectively, are deposited on the waveguide end facets, making the source doubly-resonant for the generated photons. 

This double-resonance condition modifies the joint spectral amplitude (JSA) of the generated photon pair. The resulting photon-pair component of the state can be expressed as 
\begin{equation}
	\begin{split}
	\ket{\psi}_\mathrm{RPDC} = \mathcal{B}&\int
	d\omega_\mathrm{s}d\omega_\mathrm{i}\,
	f(\omega_\mathrm{s},\omega_\mathrm{i})\times\\
	&A_\mathrm{s}(\omega_\mathrm{s})
	A_\mathrm{i}(\omega_\mathrm{i})\hat{a}^\dagger_\mathrm{s}(\omega_\mathrm{s})
	\hat{a}^\dagger_\mathrm{i}(\omega_\mathrm{i})\ket{0}.
	\end{split}
	\label{eq:pdc_state}
\end{equation}
Here, the constant $\mathcal{B}$ describes the strength of the PDC interaction, $f(\omega_\mathrm{s},\omega_\mathrm{i})=f_\mathrm{p}(\omega_\mathrm{s}+\omega_\mathrm{i})\times f_\mathrm{PM}(\omega_\mathrm{s},\omega_\mathrm{i})$ is the JSA of the non-resonant PDC, which in turn is the product of the pump spectral amplitude $f_\mathrm{p}(\omega_\mathrm{s}+\omega_\mathrm{i})$ and the phasematching function $f_\mathrm{PM}(\omega_\mathrm{s},\omega_\mathrm{i})$, $A_\mathrm{s}(\omega_\mathrm{s})$ and $A_\mathrm{i}(\omega_\mathrm{i})$ are the cavity response functions for signal and idler, and $\hat{a}^\dagger_\mathrm{s}(\omega_\mathrm{s})$ and $\hat{a}^\dagger_\mathrm{i}(\omega_\mathrm{i})$ are standard monochromatic creation operators for signal and idler photons. 

Treating the cavity as a classical Fabry-P\'erot resonator, the cavity response functions are found to be
\begin{equation}
	A_j(\omega_j)=\frac{\sqrt{(1-R_{j,1})(1-R_{j,2})} e^{-\alpha_jL/2}}
		{1 - \sqrt{R_{j,1}R_{j,2}}e^{-\alpha_jL}e^{i\phi_j(\omega_j)}},
	\label{eq:airy_function}
\end{equation}
where the round-trip phase $\phi_j(\omega_j) = 2\omega_jn(\omega_j)L/c$, $\alpha_j$ is the waveguide propagation loss and $j={\mathrm{s},\mathrm{i}}$.

\begin{figure}
	\centering
	\includegraphics[width=\linewidth]{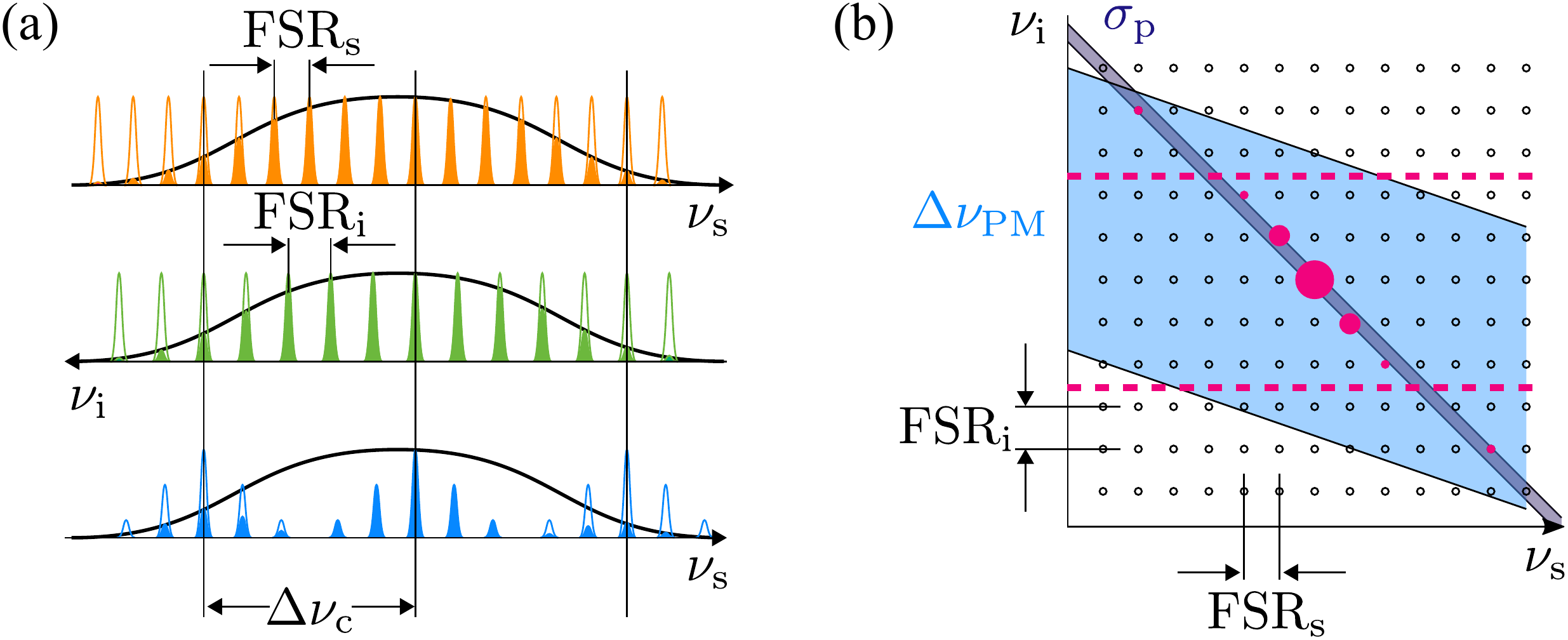}
	\caption{Schematic representation of the formation of frequency clusters in a doubly-resonant PDC. (a) The doubly-resonant cavity has dissimilar free spectral ranges for the signal (orange) and idler (green) fields. Simultaneous resonance of both fields is only possible for distinct frequency combinations, shown by vertical black lines. This leads to the formation of so-called clusters (blue). Note that the cavity structure is further modified by the phasematching of the PDC (solid black line). (b) Two-dimensional representation of the clustering in the signal and idler frequency plane. Double-resonances  require an overlap of the cavity peaks (black circles), the pump envelope function (dark blue line) and the phasematching function (light blue area), which is indicated by magenta discs. The dashed magenta lines symbolise an additional filter that transmits only the predominant central cluster. For more information, see the text.}
	\label{fig:clustering}
\end{figure}

Since the generated signal and idler photons are both frequency and polarisation non-degenerate, they also experience different waveguide dispersions. This, in turn, leads to dissimilar free spectral ranges (FSRs) for signal and idler in the doubly-resonant cavity, which fosters the formation of so-called clusters. This is sketched in Fig.~\ref{fig:clustering}(a). The orange signal resonances in the top plot overlap with the green idler resonances in the centre plot only for certain frequency pairs -- the clusters -- indicated by vertical black lines with a cluster frequency spacing $\Delta\nu_\mathrm{c}$. The resulting joint distribution is shown in blue in the bottom plot. We emphasise that the signal and idler frequency axes point in opposite directions, such that energy conservation is guaranteed for any vertical combination of frequencies.

Note that the empty peaks correspond to the cavity structure only, whereas the solid peaks are additionally modified by the phasematching function, shown as solid black line in all three plots. We have shown in \cite{Luo:2015vz}, that the FWHM of the spectrum of signal photons generated in a non-degenerate source (i.e. the bandwidth as defined by the phasematching function only) can be approximated as
\begin{equation}
	\Delta\nu_\mathrm{signal}\approx5.56\frac{c}{2\pi L}
	\frac{1}{|n_\mathrm{g,s}-n_\mathrm{g,i}|}.
\end{equation}
Here, $L$ denotes the waveguide length and $n_\mathrm{g,s}$ and $n_\mathrm{g,i}$ are the group refractive indices of signal and idler, respectively. In addition, the cluster spacing $\Delta\nu_\mathrm{c}$ is given by
\begin{equation}
	\Delta\nu_\mathrm{c}\approx\frac{c}{2L}
	\frac{1}{|n_\mathrm{g,s}-n_\mathrm{g,i}|}.
\end{equation}
These results show that the non-resonant bandwidth and the cluster spacing have a constant ratio, which is inherent to our design. When one cluster is at the centre of the phasematched signal spectrum, excitation of two weak side-clusters in the wings of the phasematching function cannot be eliminated, as shown in Fig.~\ref{fig:clustering}(a). In this case, the predominant central cluster comprises 90\% of the generated photons. Note that we will demonstrate later, how the central cluster can be separated from the other clusters and engineered such that all heralded photons are emitted into one single mode.

In Fig.~\ref{fig:clustering}(b) we sketch this situation in the two-dimensional signal and idler frequency space typically associated with the JSA. The cavity resonances are now given by the grid of black circles. The phasematching function is drawn as light blue area and characterised by the phasematching bandwidth $\Delta\nu_\mathrm{PM}$. The pump envelope function is shown as dark blue area. Due to reasons of energy conservation it is oriented along $-45^\circ$ and its width is given by the pump spectral bandwidth $\sigma_\mathrm{p}$. Note that for our design, $\Delta\nu_\mathrm{PM}$ is significantly larger than $\sigma_\mathrm{p}$. Now, only cavity resonances that overlap with both the phasematching and the pump envelope function are excited, as schematically depicted by the magenta discs, the diameter of which reflects the height of the excited peak. The dashed magenta lines denote a possible broadband filter that can be applied to the idler photons. Its bandwidth is chosen such that the parasitic clusters in the wings of the phasematching function are rejected and detection of an idler photon heralds a signal photon, which resides in the predominant central cluster. For typical source configurations, the cluster spacing $\Delta\nu_\mathrm{c}$ is on the order of several tens of GHz. In this case, the idler filtering can be realised with an off-the-shelf volume Bragg grating. We will assume this type of filtering in the remainder of this paper, to keep our notation concise. Note that in a typical quantum repeater scheme, the signal photons will intrinsically be filtered by the narrow transition bandwidth of the quantum memory. Hence, no additional signal filter is required. 

\section{Performance benchmarks associated with our PDC design}
When designing a photon-pair source that links with a quantum memory, several benchmarks are important. These are: the central frequency and spectral bandwidth of the photons that interact with the quantum memory; the photon purity; the brightness of the source, that is the normalised number of generated photon pairs; and the capability to fine-tune the source emission. In the following, we study how fabrication and operation parameters influence these benchmarks for our source design. In particular, we investigate the reflectivities of the dielectric coatings, the waveguide temperature and length, as well as the pump spectral bandwidth and central frequency. We will assume that the poling period $\Lambda$ is chosen to facilitate perfect phasematching for the desired pump, signal and idler frequencies for a given sample temperature. 
\subsection{Photon spectral bandwidth}
An important pair of benchmarks when designing a PDC source for hybrid quantum networks is the central frequency and spectral bandwidth of the generated photons. By convention, we require the signal photon to match the desired transition of the quantum memory. In an ideal resonant source with only one excited mode, the spectral bandwidth of the signal photon equals the width of the signal resonances. We will investigate this condition for realistic sources in more detail later. The bandwidth of the resonances $\Delta\nu$ can be calculated from the finesse of the cavity $\mathcal{F}$ and the corresponding FSR, since
\begin{equation}
	\mathcal{F} = \frac{\mathrm{FSR}}{\Delta\nu}.
\end{equation}
To study the influence of the waveguide length $L$ and the reflectivity of the second dielectric coating $R_\mathrm{s,2}$, we consider the wavelength combination of our demonstrator source \cite{Luo:2015vz}. When pumped with $532\,$nm, it generates signal and idler photons at $890\,$nm and $1320\,$nm, respectively.  The actual demonstrator was realised in a $12.3\,$mm long waveguide with propagation losses of $\alpha_\mathrm{s}=0.016\frac{\mathrm{dB}}{\mathrm{cm}}$ and $\alpha_\mathrm{i}=0.022\frac{\mathrm{dB}}{\mathrm{cm}}$. The endfacet coatings had reflectivities of $R_\mathrm{s,1}=R_\mathrm{i,1}=0.99$ and $R_\mathrm{s,2}=R_\mathrm{i,2}=0.98$, respectively, yielding a signal spectral bandwidth of around $66\,$MHz. 

\begin{figure}
	\centering
	\includegraphics[width=\linewidth]{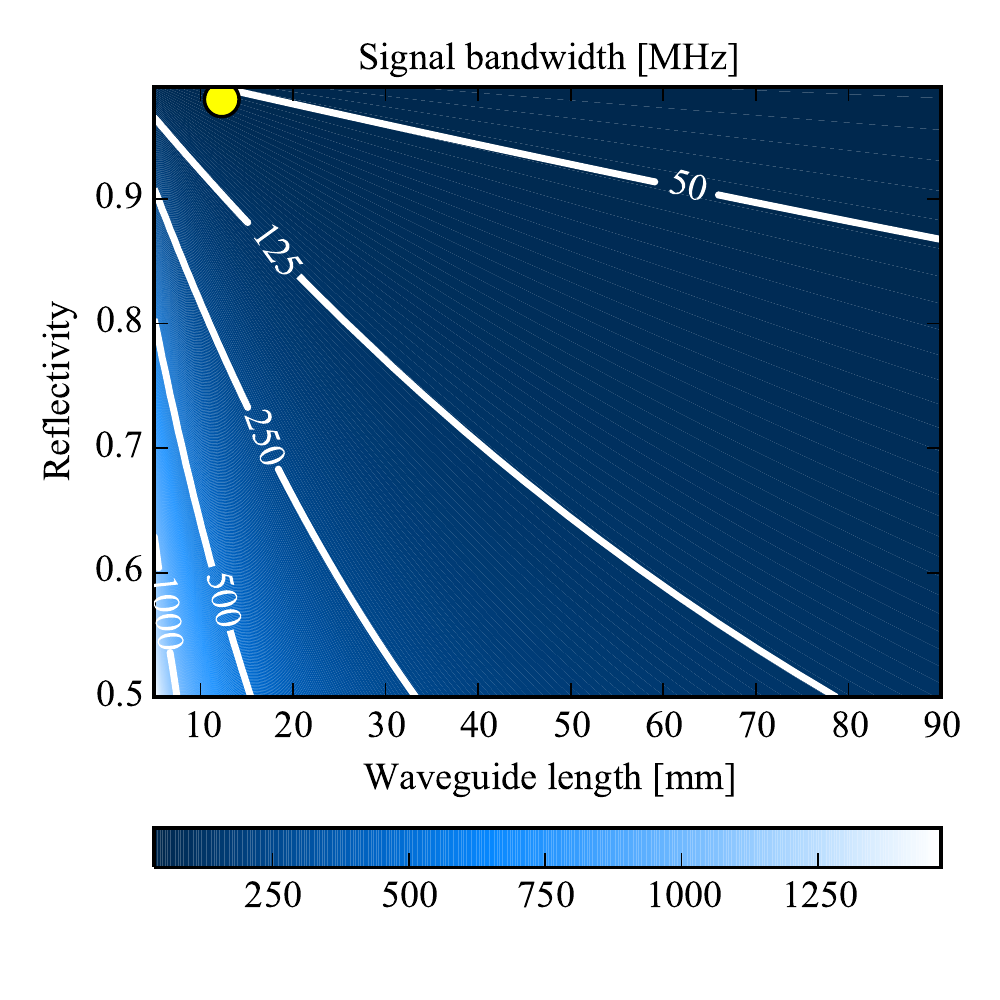}
	\caption{Calculated signal resonance spectral bandwidth as a function of waveguide length and reflectivity of the second dielectric coating. The yellow disc is our demonstrator source. Accessible bandwidth range from a few tens of MHz to around one GHz and each bandwidth can be realised with different combinations of waveguide length and coating reflectivity.}
	\label{fig:bandwidth}
\end{figure}

In Fig. \ref{fig:bandwidth} we plot the cavity-peak bandwidth for the generated signal as function of $L$ and $R_\mathrm{s,2}$, while keeping the other parameters at the aforementioned values. The yellow circle denotes our demonstrator. We can draw several conclusions from this plot. On the one hand, broad bandwidths in excess of $1\,$GHz can only be achieved with very short waveguides with lengths below $5\,$mm. On the other hand, small bandwidths of tens of MHz require very long waveguides, and the minimum bandwidth is limited by waveguide propagation losses. 

Still, the realisable bandwidths are compatible with many quantum memories, for instance rare-earth doped solid state memories or Raman memories \cite{Simon:2010kl,Bussieres:2013br}. In addition, a specific bandwidth can be realised for a wide range of waveguide lengths, if the reflective coating is properly adapted. This loosens the requirements for waveguide fabrication tolerances. 
\subsection{Photon central frequency}
The second benchmark we study in this section is the central frequency of the generated signal photon, which depends on both the pump central frequency and the sample temperature. Let us, for the moment, assume an ideal cw pump laser. Then, the signal photon spectrum can be expressed as \cite{Luo:2015vz}
\begin{equation}
	S(\omega_\mathrm{s}) \propto
	\operatorname{sinc}^2\left(\Delta\beta(\omega_\mathrm{s})
	\frac{L}{2}\right)|A_\mathrm{s}(\omega_\mathrm{s})|^2\,
	|A_\mathrm{i}(\omega_\mathrm{p}-\omega_\mathrm{s})|^2,
\end{equation}
where $\Delta\beta(\omega_\mathrm{s})$ is the phase-mismatch (compare Eq.~(\ref{eq:phasematching})) and the idler frequency is calculated using the energy conservation in Eq.~(\ref{eq:energy}). 

\begin{figure}
	\centering
	\includegraphics[width=\linewidth]{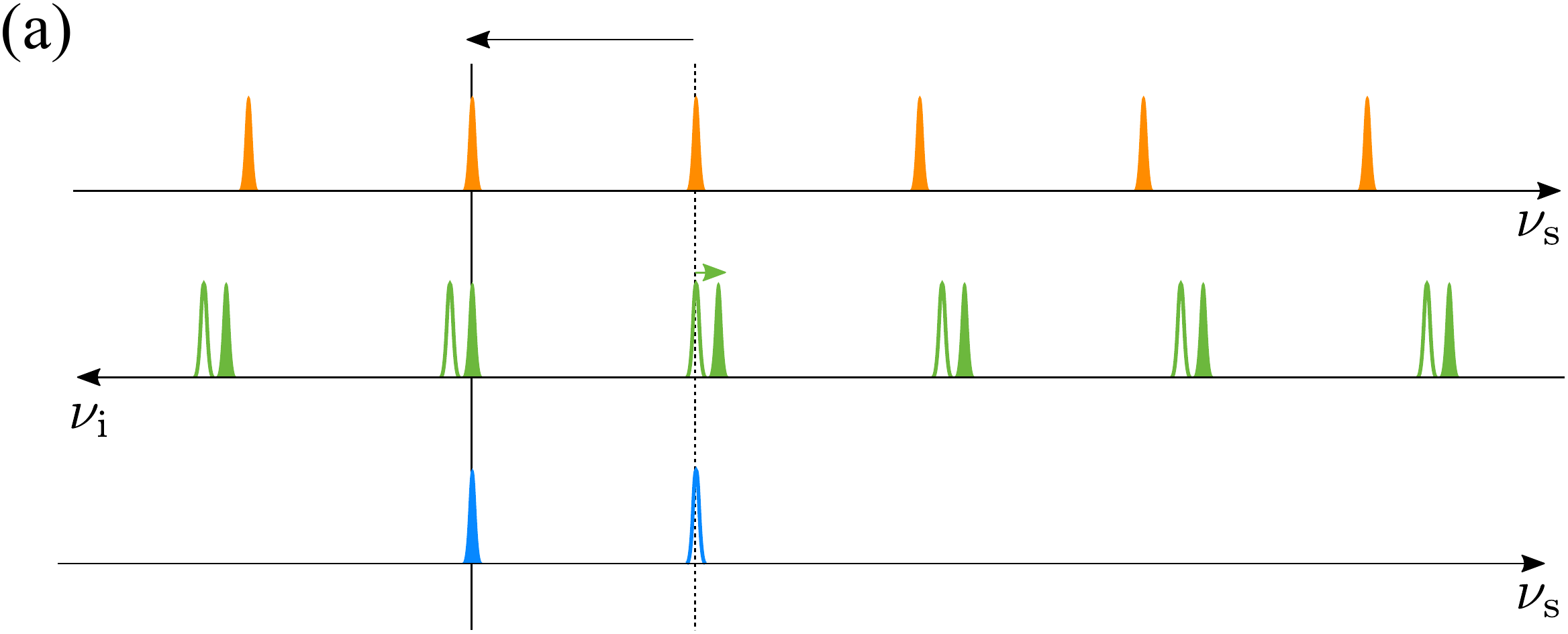}
	\includegraphics[width=\linewidth]{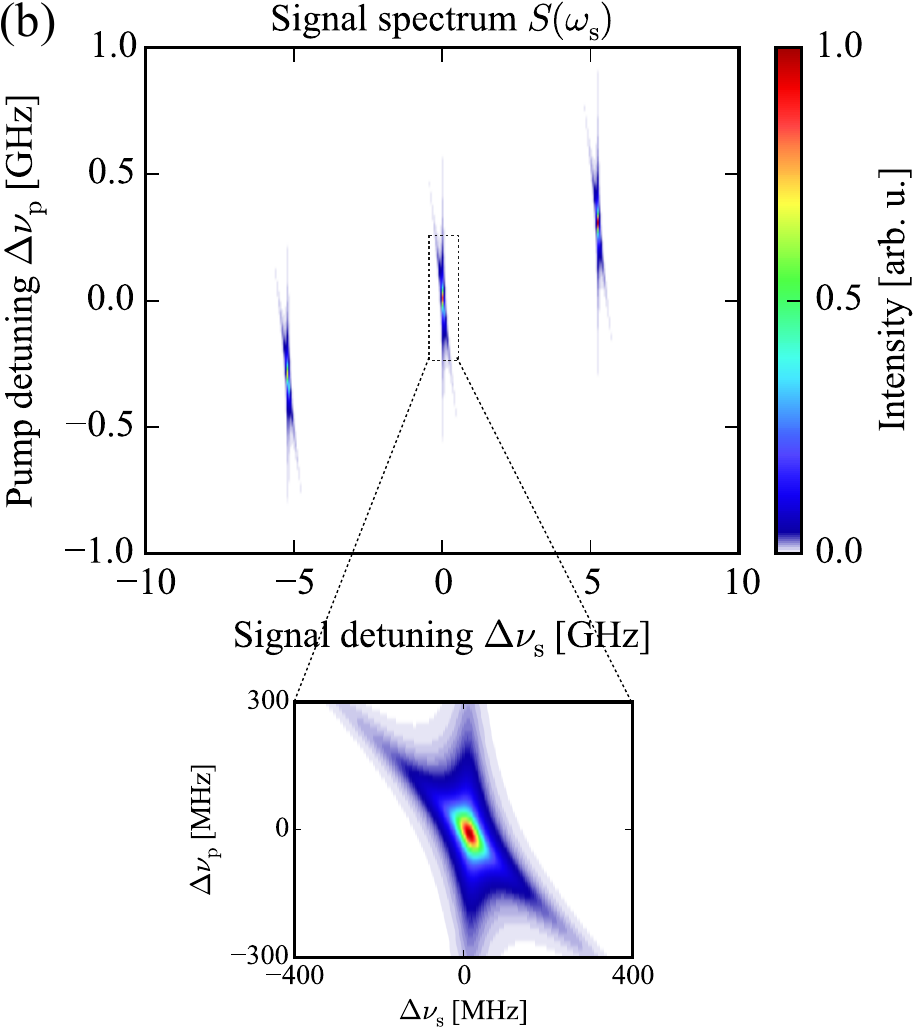}
	\caption{Effect of pump frequency tuning on the signal photon central frequency. (a) When tuning the pump frequency, energy conservation dictates that the idler resonances (green) are shifted with respect to the signal resonances (orange). If the pump shift is sufficiently large, a mode hop occurs in the signal spectrum (blue). (b) Signal spectrum $S(\omega_\mathrm{s})$ as function of the pump frequency detuning. As expected, mode hops between adjacent signal resonances occur for a pump detuning larger than around $300\,$MHz. Faithful single-mode operation requires a pump frequency stability of $\pm100\,$MHz. The zoom-in into one resonance reveals an internal structure that stems from the interplay between signal and idler resonances. For more details, see the text.}
	\label{fig:pump_tuning}
\end{figure}

In Fig.~\ref{fig:pump_tuning}(a), we schematically depict the influence of pump frequency tuning. Energy conservation dictates that $\nu_\mathrm{p} = \nu_\mathrm{s} + \nu_\mathrm{i}$. Then, a change in the pump frequency is compensated by a joint shift in signal and idler frequencies. In our graphic interpretation, this is shown by a common shift of all idler resonances (green peaks) relative to the signal resonances (orange peaks). If the shift is large enough, this results in a mode hop in the signal spectrum (blue peaks). Note that the mode hops described here aren't similar to the typical hopping behaviour in a laser cavity, which arises due to gain competition between different longitudinal resonator modes. Rather they are caused by the PDC emission passively following the phasematching function.

In Fig.~\ref{fig:pump_tuning}(b) we plot the signal spectrum $S(\omega_\mathrm{s})$ as a function of the pump detuning $\Delta\nu_\mathrm{p}$ for a fixed sample temperature of $T=148.14^\circ\,$C. The x-axis is given in terms of signal detuning $\Delta\nu_\mathrm{s}$. As expected, we observe a mode hop in the signal spectrum if the pump detuning is too large. The inset shows a zoom into one of the resonances. The cross-shaped structure arises from the fact that, by tuning the pump frequency, the idler resonances shift with respect to the signal resonances. Requiring a mode-hop free source operation, we can infer a stability criterion for the pump frequency. For the case considered here, it evaluates to $|\Delta\nu_\mathrm{p}|\lesssim100\,$MHz. This can be satisfied with commonplace external cavity lasers and frequency locking techniques.

\begin{figure}
	\centering
	\includegraphics[width=\linewidth]{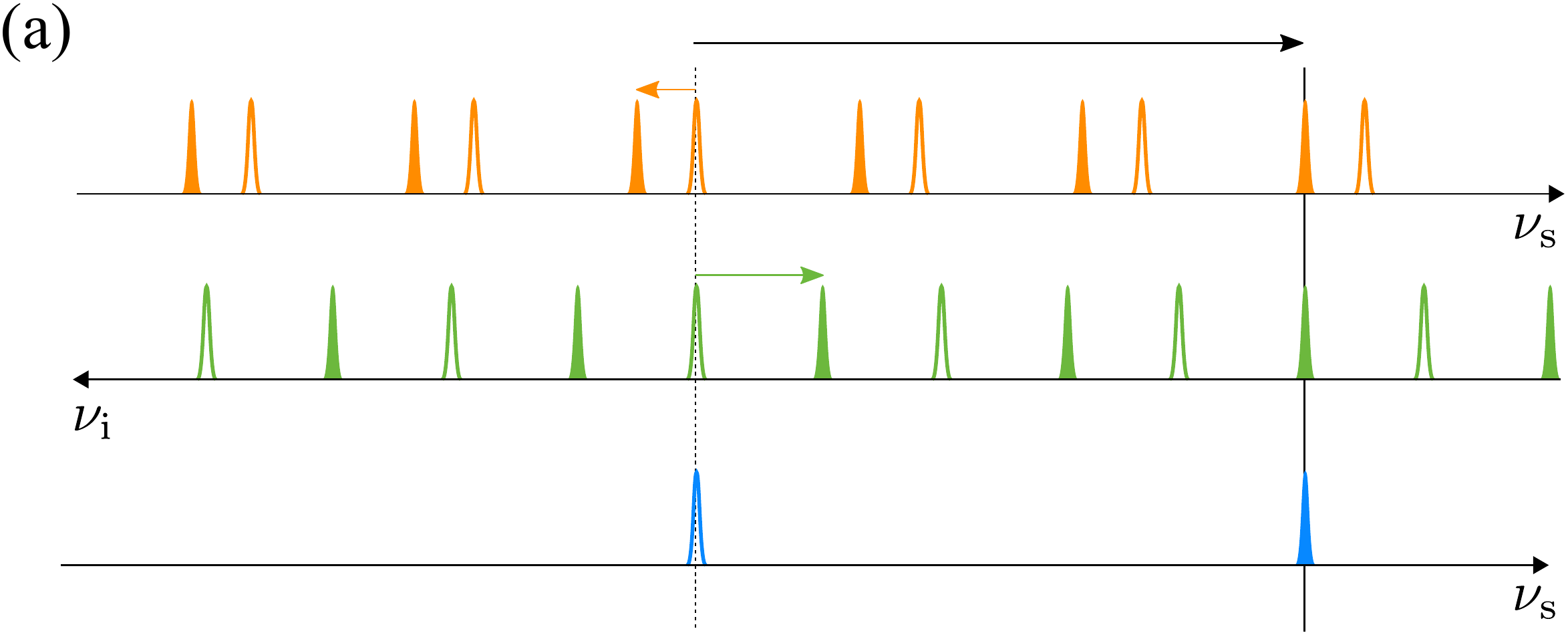}
	\includegraphics[width=\linewidth]{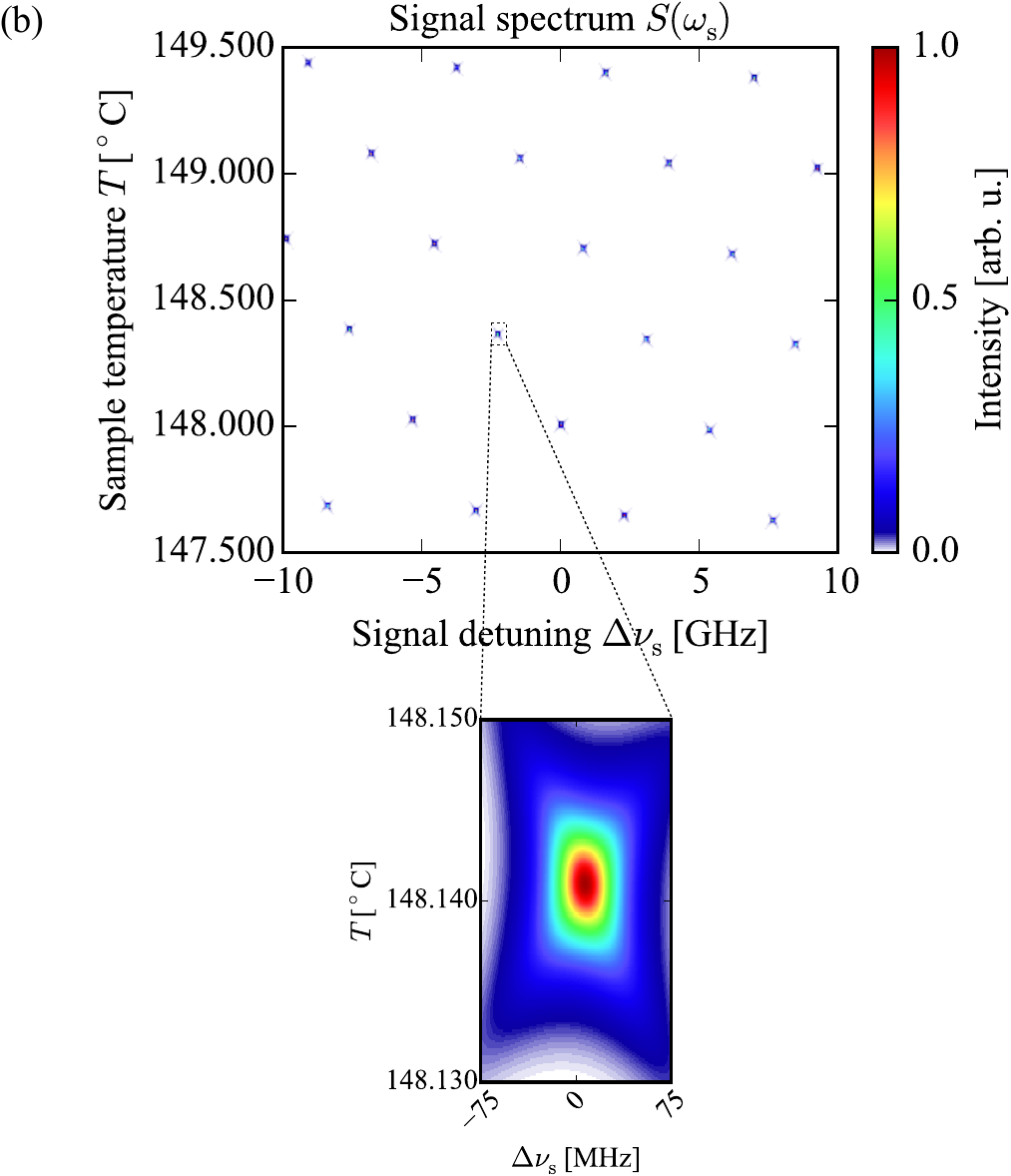}
	\caption{Effect of temperature tuning on the signal photon central frequency. (a) The signal (orange) and idler (green) resonances shift in opposite directions and at different velocity when the sample temperature changes. For large temperature drifts, this can lead to large mode hops in the signal spectrum (blue). (b) Signal spectrum $S(\omega)$ as function of the sample temperature $T$. As expected, mode hops occur mainly between adjacent modes for small temperature drifts. Notably, the absolute position of the signal peaks shifts to larger frequencies for increasing temperatures. This facilitates coarse frequency tuning by adapting the sample temperature. Again, the zoom-in shows a single resonance, from which we deduce a temperature stability requirement of $\pm5\,$mK for single-mode operation.}
	\label{fig:temperature_tuning}
\end{figure}

Now, we turn our attention to the required temperature stability. In Fig.~\ref{fig:temperature_tuning}(a), we sketch the behaviour of the signal and idler resonances when changing the sample temperature. A shift in temperature causes two effects: first, the physical sample length changes due to thermal expansion; second, the optical path length changes due to a change in the temperature-dependent refractive index. We note that the thermal expansion for a temperature shift of only $10\,$mK is on the order of $1.5\,$nm per cm sample length and thus cannot be neglected. As a result of these changes, signal (orange peaks) and idler (green peaks) resonances shift in opposite directions with different speed, highlighted by the coloured arrows, and large mode hops can occur in the signal spectrum (blue peaks) for large enough temperature shifts. 

Fig.~\ref{fig:temperature_tuning}(b) shows $S(\omega_\mathrm{s})$ as function of the sample temperature, for a fixed pump wavelength of $532\,$nm. Small temperature changes mostly cause mode hopping to neighbouring resonances, until at some point a larger hop occurs. In addition, the position of the signal resonances gradually moves towards larger frequencies with increasing temperature. We will come back to this behaviour later, when we discuss frequency fine-tuning of our source. Again, the inset shows a zoom-in into one signal resonance, and we deduce a stability region of around $\pm5\,$mK around the ideal temperature setpoint, which guarantees mode-hop free operation. 

We find that the stability requirements on pump frequency and sample temperature can be satisfied with off-the-shelf components, which highlights the practicality of our design for real-world applications. 
\subsection{Photon purity}
In order for the generated photons from our source to be useful in large-scale hybrid quantum networks, we require them to be pure. Let us define the term \textit{purity} for our design in a two-step approach. First, we consider how much of the heralded single photon resides in the dominant resonator mode. We call this quantity the \textit{mode-excitation probability} $\mathcal{M}$. Then, we investigate the \textit{spectral purity} $\mathcal{S}$ of the part of the photon inside the dominant resonator mode. Finally, we define a total purity $\mathcal{P}=\mathcal{M}\cdot\mathcal{S}$ as the product of the two partial quantities. A PDC source that is ideally adapted to quantum memories generates spectrally pure single photons in only a single resonator mode, and thus features $\mathcal{P}=1$. Note that we do not consider photon-number purity here. However, this value can be made high by using low pump powers to reduce the probability of multi-photon-pair generation events, and by optimisation of the optical setup to reduce heralding losses. 

\begin{figure}
	\centering
	\includegraphics[width=\linewidth]{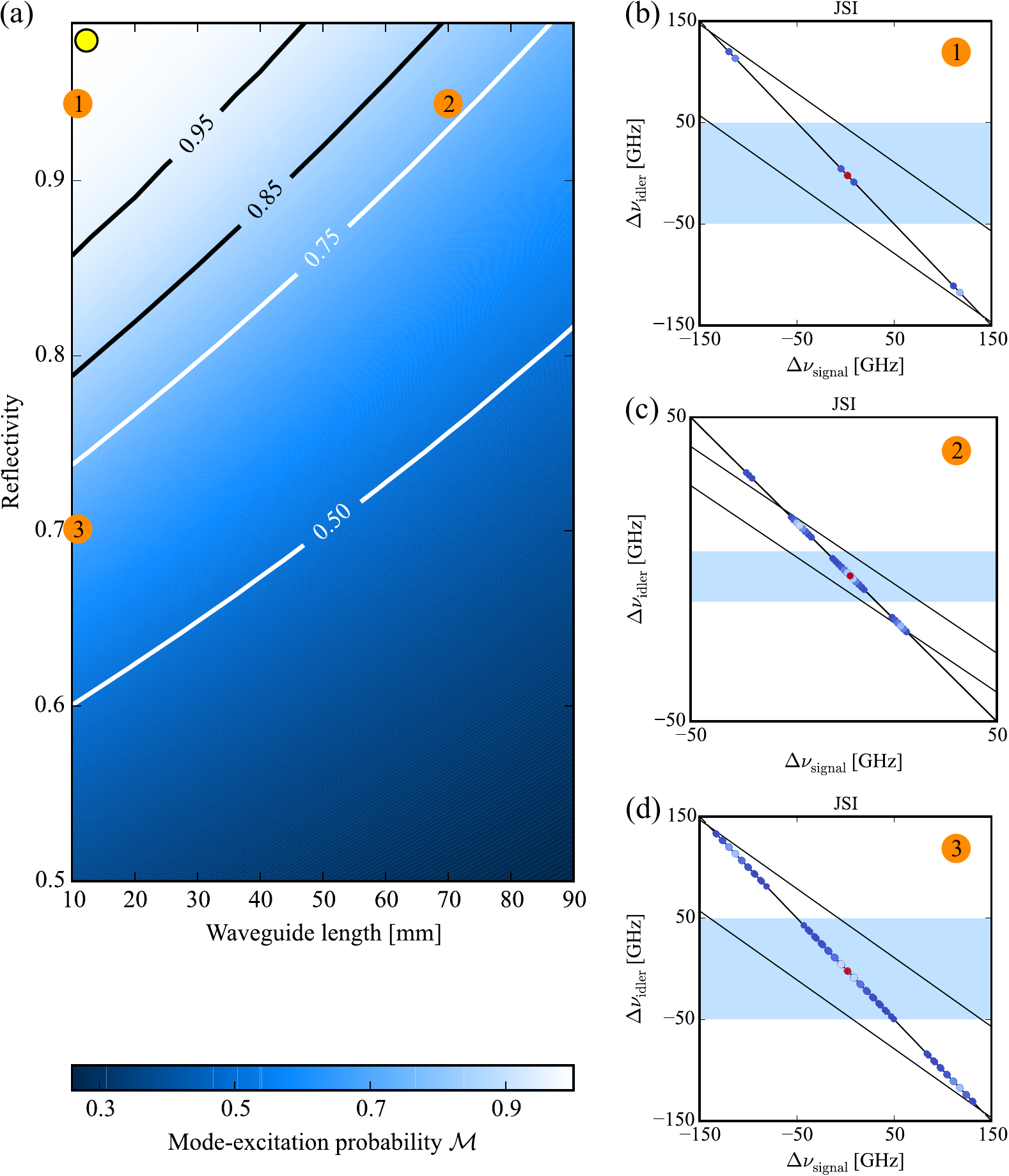}
	\caption{Mode-excitation probability $\mathcal{M}$ as a function of waveguide length and reflectivity of the second endfacet coating. (a) The plot shows a clear trend towards short samples and high reflectivities, with our demonstrator sitting in the upper left corner (yellow disc). This can be understood when considering specific combinations of length and reflectivity, marked with numbers 1-3. (b) The JSI of a doubly-resonant PDC with a waveguide length of $10\,$mm and a reflectivity of 95\%. Due to the narrow linewidth of the resonances and their large separation, mainly a single resonance is excited, leading to a high $\mathcal{M}$. (c) The JSI for a $70\,$mm long waveguide with a reflectivity of 95\%. The reduced distance between adjacent resonances leads to more resonances being excited. Consequently, $\mathcal{M}$ is lower than for (b). (d) The JSI for a $10\,$mm long waveguide with 70\% reflectivity. Although the resonances are far apart, the low reflectivity leads to a large bandwidth of the single resonances. As a consequence multiple resonances are excited and, again, we find a low $\mathcal{M}$. For detailed explanations, see the text.}
	\label{fig:mode_excitation}
\end{figure}

First, we consider the mode-excitation probability $\mathcal{M}$, which is a measure for the percentage of \textit{useful} heralded photons, that is photons whose frequency is matched to the desired memory transition. Fig.~\ref{fig:mode_excitation}(a) shows $\mathcal{M}$ as a function of waveguide length $L$ and reflectivity of the second dielectric coating $R_\mathrm{s,2}$. Again, the yellow circle denotes our demonstrator from \cite{Luo:2015vz}. Fig.~\ref{fig:mode_excitation}(b)-(d) show the joint spectral intensity (JSI) distributions for selected combinations of $L$ and $R_\mathrm{s,2}$. The parallel black lines denote the phasematching function and the additional line is the pump distribution (compare Fig.~\ref{fig:clustering}). The blue-shaded area shows the applied spectral filter, which transmits only the central cluster. The JSI is the modulus squared of the modified JSA function. For a short $L$ and high $R_\mathrm{s,2}$ as shown in (b), only few cavity modes are excited as they are well separated in the signal and idler frequency space. When increasing $L$ while keeping $R_\mathrm{s,2}$ constant, the FSR between the cavity modes decreases and more modes are excited, resulting in a smaller $\mathcal{M}$. The corresponding JSI is plotted in (c). A similar behaviour is found for constant $L$ and decreasing $R_\mathrm{s,2}$. In this case, however, the FSR stays constant and the spectral bandwidth of the cavity modes increases, as shown in (d). Again, this results in more modes being excited and thus a decrease in $\mathcal{M}$. Generally, we find that values of $\mathcal{M}>0.95$ can be achieved for numerous combinations of $L$ and $R_\mathrm{s,2}$, corresponding to generated photon spectral bandwidths of up to around $350\,$MHz (compare Fig.~\ref{fig:bandwidth}). 

\begin{figure}
	\centering
	\includegraphics[width=\linewidth]{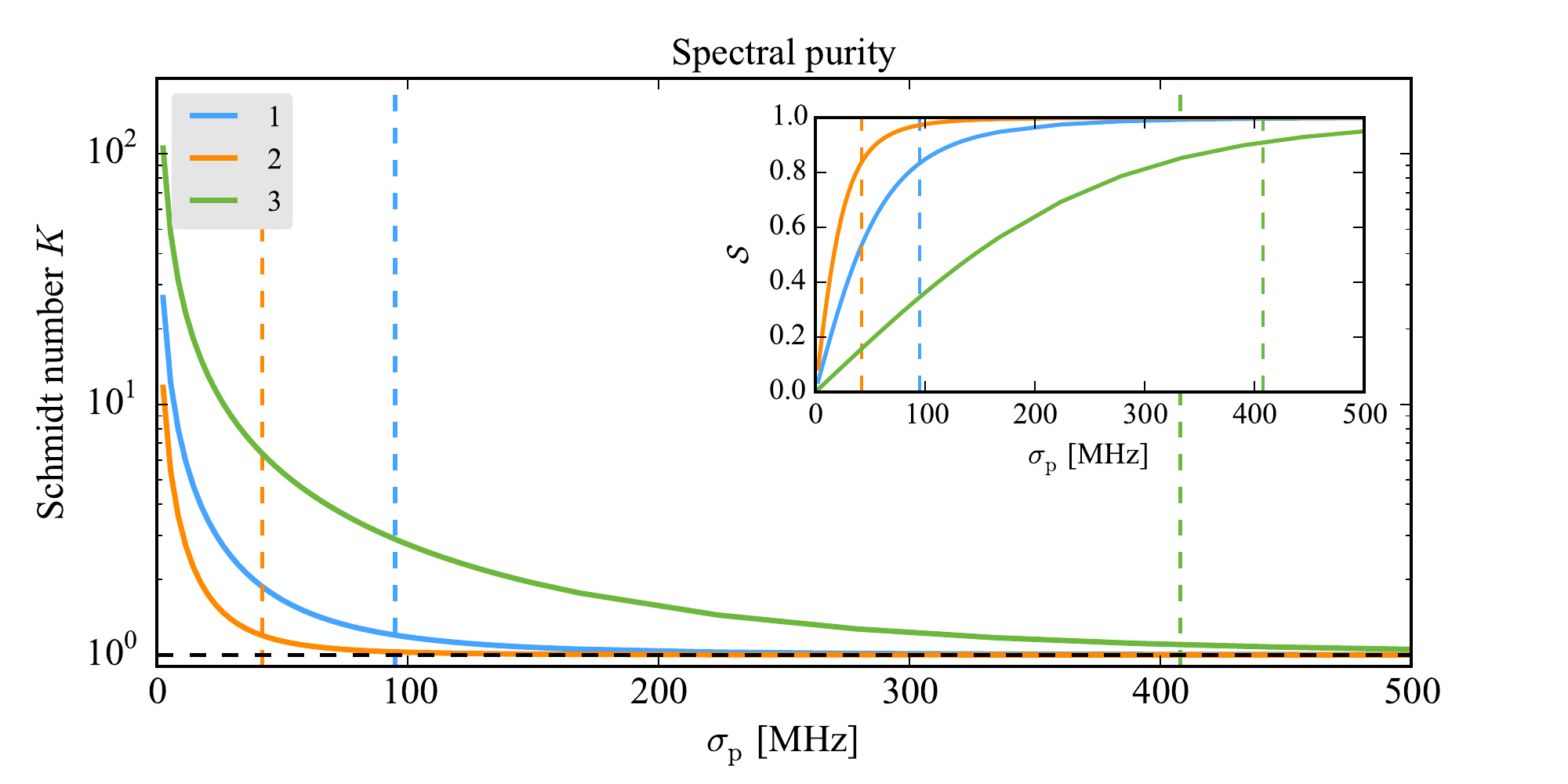}
	\caption{Photon spectral purity. A Schmidt decomposition of the JSI of the most dominant resonance yields the Schmidt number $K$ as a function of pump spectral bandwidth $\sigma_\mathrm{p}$, which is a measure for the effective number of spectral modes of the signal. The labelling corresponds to the chosen combinations of waveguide length and coating reflectivity from Fig.~\ref{fig:mode_excitation}. The vertical dashed lines denote the corresponding spectral bandwidths of the resonances. A Schmidt number close to one is obtained for cases when $\sigma_\mathrm{p}$ is roughly twice the resonance bandwidth or larger. The inset shows the spectral purity of the photons, which can be calculated from the Schmidt number as $\mathcal{S}=1/K$. High spectral purities require pump spectral bandwidths that are large compared to the resonance bandwidth.}
	\label{fig:spectral_purity}
\end{figure}

Now, we investigate the spectral purity of the heralded photons. To this end, we perform a Schmidt decomposition of the JSI of the most dominant cavity mode \cite{Law:2000wd}. From this, we retrieve the Schmidt number $K$, which is a measure for the effective number of spectral modes \cite{URen:2005wb}. It is also the inverse of the spectral purity $\mathcal{S}=1/K$. Note that de-composing the JSI only provides a lower bound for $K$, since possible correlations in the phase of the JSA are not captured \cite{Brecht:2013fq}. 

We plot the Schmidt number as a function of the pump spectral bandwidth $\sigma_\mathrm{p}$ in Fig.~\ref{fig:spectral_purity}.  Note that this implicitly assumes a pulsed pump laser with pulse durations of several hundreds of ps to several ns, while our demonstrator source was driven with a cw laser. The labelling corresponds to the selected combinations of $L$ and $R_\mathrm{s,2}$ from Fig.~\ref{fig:mode_excitation}(a). The dashed vertical lines denote the spectral bandwidth of the resonances. The inset shows the spectral purity $\mathcal{S}$ as derived from the Schmidt number. As long as the pump spectral bandwidth is smaller than roughly twice the resonance bandwidth, the pump induces spectral correlations between the two generated photons. This results in a spectrally impure heralded single photon. In contrast, as soon as the pump is broader than twice the resonance bandwidth, the PDC photons become spectrally decorrelated and the detection of an idler heralds a spectrally pure signal photon.

With these results, we can make a statement on the overall purity $\mathcal{P}$ of the heralded photon for a given set of parameters. For the three selected pairs of $L$ and $R_\mathrm{s,2}$ and an arbitrarily chosen pump spectral bandwidth of $\sigma_\mathrm{p}=100\,$MHz, we find purities of $\mathcal{P}_1\approx0.81$, $\mathcal{P}_2\approx0.72$ and $\mathcal{P}_3\approx0.24$, respectively. For our demonstrator source, we find $\mathcal{P}\approx0.95$ for $\sigma_\mathrm{p}\approx100\,$MHz. However, we note that in our measurements we were pumping with a narrowband pump, which reduces the purity to $\mathcal{P}\approx0.50$. 

The calculations demonstrate that our source design is capable of generating high-purity photons for a large diversity of design parameters. Notably, high spectral purity photons can be generated over the full range of available spectral bandwidths. For the extreme cases of very narrow or broad bandwidths, this leads, however, to a reduction in mode-excitation probabilities. Note that this is not necessarily a prohibitive criterion for quantum repeater applications, where repeat-until-success protocols can be utilised. Then, a low mode-excitation probability translates to a larger number of required experiments to establish success. 
\subsection{Source brightness}
The doubly-resonant design of our source enhances the photon generation rate by a factor that is proportional to the product of signal and idler finesse $\mathcal{F}_\mathrm{s}\cdot\mathcal{F}_\mathrm{i}$ \cite{Ou:1999wg,JeronimoMoreno:2010kq} and an additional clustering factor $N_0$ \cite{Luo:2015vz}. A large finesse facilitates a high enhancement of the PDC, but also means that the generated photons experience more cavity roundtrips and thus higher losses. This reduces the effective enhancement; the reduction can be quantified by introducing the so-called photon-pair escape probability $\eta_\mathrm{pp}$ \cite{Luo:2015vz}. Thus, we find a total enhancement that is proportional to
\begin{equation}
	M\propto 
	N_0\mathcal{F}_\mathrm{s}\,\mathcal{F}_\mathrm{i}\,\eta_\mathrm{pp},
\end{equation}
where
\begin{align}
	N_0 &= 2n_\mathrm{g,s}L\frac{\Delta\nu_\mathrm{c}}{c},\\
	\mathcal{F}_\mathrm{s/i}&=\frac{\pi\sqrt{\sqrt{R_\mathrm{s/i,1}R_\mathrm{s/i,2}}
	e^{-\alpha_\mathrm{s/i}L}}}{1 - \sqrt{R_\mathrm{s/i,1}R_\mathrm{s/i,2}}
	e^{-\alpha_\mathrm{s/i}L}},\\
	\eta_\mathrm{pp}&=\frac{1-R_\mathrm{s,2}}{1-R_\mathrm{s,1}R_\mathrm{s,2}
	e^{-2\alpha_\mathrm{s}L}}\times\frac{1-R_\mathrm{i,2}}{1-R_\mathrm{i,1}
	R_\mathrm{i,2}e^{-2\alpha_\mathrm{i}L}}.
\end{align}  

In addition, the spectral density of the generated PDC photons is redistributed by the cavity, leading to more photons being generated in the spectral region of interest (the memory transition).  As soon as the length, loss and mirror coatings of a resonant waveguide are fixed, the photon generation efficiency can be estimated with these equations. For our demonstrator we inferred a spectral brightness of $B\approx3\cdot10^4\,\frac{1}{\mathrm{s}\,\mathrm{mW}\,\mathrm{MHz}}$. This means that the total generation rate in the $66\,$MHz
 wide resonance evaluates to $2\cdot10^6\,\frac{1}{\mathrm{s}\,\mathrm{mW}}$. From the measured generation rate of a non-resonant guided-wave PDC (with similar properties as our demonstrator device) and its spectral bandwidth of around $166\,$GHz, we can estimate the brightness of the non-resonant source to be typically around $10\dots20\,\frac{1}{\mathrm{s}\,\mathrm{mW}\,\mathrm{MHz}}$. Thus, the resonant source provides a brightness which is more than three orders of magnitude larger. 

In conclusion, our doubly-resonant design facilitates a mote than 1000-fold increase in spectral source brightness at the memory transition by, on the one hand, enhancing the overall photon generation rate and, on the other hand, redistributing the photon spectrum. Since most of the photons are generated at the memory transition, the energy efficiency of our source, that is the number of useful photons per unit of pump power, is very high. This is an important benchmark for large-scale quantum networks, where many devices have to be operated in parallel. 
\subsection{Frequency fine tuning}
To adapt our PDC source to a specific quantum memory transition, fine tuning of the emission frequency to match exactly the transition line of the memory is required. This is facilitated by synchronously tuning the temperature and the pump frequency.

If we assume that our cavity is resonant at a given temperature $T_0$ for both a signal frequency $\Omega_\mathrm{s}$ and an idler frequency $\Omega_\mathrm{i}$, then the round-trip phases for both waves are integer multiples of $2\pi$, that is
\begin{equation}
	\phi_\mathrm{s}(\Omega_\mathrm{s},T_0) = M\times 2\pi
	\quad\text{and}\quad
	\phi_\mathrm{s}(\Omega_\mathrm{i},T_0) = N\times2\pi,
\end{equation}
with $N$ and $M$ being integers. Note that we explicitly included the temperature dependence into the roundtrip phase $\phi$ (compare Eq.~(\ref{eq:airy_function})). 

The variation of the phase as a function of frequency and temperature detuning can be approximated by a first-order Taylor expansion according to
\begin{equation}
	\begin{split}
	\phi&=\phi(\Omega,T_0)+\partial_\omega\phi\big|_{\Omega,T_0}\Delta\omega+
	\partial_T\phi\big|_{\Omega,T_0}\Delta T\\
	&=:\phi_0+\Delta\phi.
	\end{split}
\end{equation}
To stay resonant, we require $\Delta\phi\stackrel{!}{=}0$. Thus, when tuning the signal frequency offset, the temperature must be changed accordingly such that
\begin{equation}
	\Delta T = -\frac{\partial_\omega\phi_\mathrm{s}}
	{\partial_T\phi_\mathrm{s}}\Delta\omega_\mathrm{s}.
	\label{eq:shift_01}
\end{equation}
However, this temperature change drives a frequency shift of the idler. given by
\begin{equation}
	\Delta\omega_\mathrm{i} = -\frac{\partial_T\phi_\mathrm{i}}
	{\partial_\omega\phi_\mathrm{i}}\Delta T.
	\label{eq:shift_02}
\end{equation}
The shift of both signal and idler frequencies is the required shift in pump frequency, since $\Delta\omega_\mathrm{p}=\Delta\omega_\mathrm{s}+\Delta\omega_\mathrm{i}$ due to energy conservation. From Eqs.~(\ref{eq:shift_01}) and (\ref{eq:shift_02}), it follows that
\begin{equation}
	\Delta\omega_\mathrm{p}=\left(1 + \frac{\partial_T\phi_\mathrm{i}
	\partial_{\omega_\mathrm{s}}\phi_\mathrm{s}}
	{\partial_{\omega_\mathrm{i}}\phi_\mathrm{i}
	\partial_T\phi_\mathrm{s}}\right)
	\Delta\omega_\mathrm{s}.
\end{equation}
From the well-known dispersion characteristics of our waveguides \cite{Strake:1988ui} and the temperature dependence of the refractive indices, we derive conditions for smoothly tuning the signal frequency by adapting simultaneously the sample temperature and pump frequency:
\begin{equation}
	\Delta T\approx-0.157\frac{^\circ\mathrm{C}}{\mathrm{GHz}}
	\Delta\nu_\mathrm{s}
	\quad\text{and}\quad
	\Delta\nu_\mathrm{p}\approx2.429\Delta\nu_\mathrm{s}.
\end{equation}

\begin{figure}
	\centering
	\includegraphics[width=\linewidth]{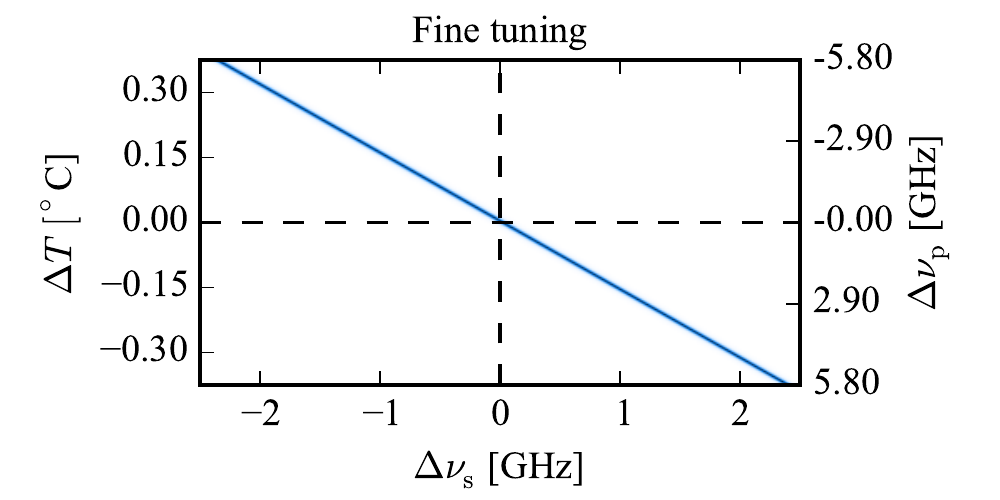}
	\caption{Frequency fine tuning of the PDC emission. The plot shows the predominant signal resonance position as a function of simultaneous temperature and pump frequency offset. Signal tuning over roughly one FSR can be achieved without loosing the resonance condition. Thus, the PDC emission can be adapted to the exact memory transmission line.}
	\label{fig:fine_tuning}
\end{figure}
We show this in Fig.~\ref{fig:fine_tuning}, where we plot the central signal resonance as it is fine-tuned over a range of $-2.3\,\mathrm{GHZ}\leq\Delta\nu_\mathrm{s}\leq2.3\,\mathrm{GHz}$. The two y-axes denote the required temperature detuning $\Delta T$ from the ideal temperature of $T(\Delta\omega_\nu{s}=0)=148.14^\circ$C, as well as the required pump frequency offset, as calculated above. Clearly, single-mode operation is retained over the full tuning range. We note that state-of-the-art temperature controllers with stabilities in the mK regime and stable lasers facilitate a reproducible tuning of our source with a precision on the order MHz, thus allowing to faithfully adapt the PDC to a given memory transition. 
\section{Adaptability}
To this point, we have focussed on our demonstrator source. However, our scheme is more flexible than this, in that it facilitates the addressing of diverse quantum memories based on different material systems. In the following, we want to name but a few examples of how the design parameters have to be adjusted to suit different experimental situations. 

\textit{Caesium vapour -- }Raman interactions in atomic vapours can be exploited to realise quantum memories, by transferring an atomic excitation from a ground state to a stable storage state. One example for this has been presented in \cite{Reim:2011gr}, where light at a wavelength of $852\,$nm is stored inside a warm vapour of Cs atoms. The bandwidth of this memory is in principle only limited by the hyperfine groundstate splitting of $9.2\,$GHz between the ground and storage states. Adapting our source to this memory and aiming for a photon bandwidth of $500\,$MHz, a pump bandwidth $\sigma_\mathrm{p}=1\,$GHz and a reflectivity $R_\mathrm{s,2}=0.90$ at a waveguide length of $L=2.5\,$mm facilitate the generation of single photons with an overall purity of $\mathcal{P}=0.940$. Note that this value, as well as the following values, are calculated for the case where the central cluster is filtered (compare Fig.~\ref{fig:mode_excitation}(b)). The corresponding idler photons are centred at $1416\,$nm, where they can be detected with state-of-the-art single photon detectors. 

\textit{Thulium ions -- }Rare-earth doped solids can be deployed as quantum memories with long coherence time and intrinsic multimode capability (see, for instance, \cite{Longdell:2005ik,deRiedmatten:2008ck,Afzelius:2009gc,Afzelius:2010fh,Clausen:2011bw,Saglamyurek:2011js,Heinze:2013dh,Jobez:2015gt}). Thulium ions feature a transition at $795\,$nm. Designing our source to emit signal photons at this wavelength places the associated idler photons at $1608\,$nm. Note that this wavelength can, in principle, be tuned by choosing a different pump wavelength. However, for reasons of consistency and due to the high availability of good lasers, we fix the pump wavelength at $532\,$nm here. Deploying a $10\,$mm long waveguide with $R_\mathrm{s,2}=0.94$ and a $250\,$MHz wide pump, our source is expected to produce signal photons with an overall purity of $\mathcal{P}=0.963$. 

\textit{Erbium ions -- }Finally, we consider erbium ions, which feature a transition at $1536\,$nm. In this case, the role of signal and idler photons are exchanged and the signal photons are used as heralds. For this case study, we aim for a narrow bandwidth of $50\,$MHz at $1536\,$nm, pairing the idler with a signal at $814\,$nm. The narrow bandwidth enforces a trade-off between the mode-excitation probability $\mathcal{M}$, which typically decreases with increasing pump bandwidth, and the spectral purity $\mathcal{S}$, which increases with increasing pump bandwidth (see Fig.~\ref{fig:spectral_purity}). We find a decent overall purity of $\mathcal{P}=0.753$ for a waveguide length of $L=80\,$mm, $R_\mathrm{s,2}=0.98$ and a pump bandwidth of $\sigma_\mathrm{p}=60\,$MHz. Although this value is somewhat lower than the purities presented before, it still means that 75\% of the generated photons are useful photons that can couple to the quantum memory. 
\section{Conclusion}
We presented an in-depth study of a flexible and versatile source design for a PDC source that generates photons suited for interaction with different quantum memories. Our source design is based on doubly-resonant PDC in a nonlinear lithium niobate waveguide and is extraordinarily robust and compact, owing to its monolithically integrated structure. Thus, it is well-suited for use in real-world applications and lends itself to integration with existing single-mode fibre architectures. 

Our investigations show, that our source design can be deployed to generate photons with spectral bandwidths ranging from few tens of MHz up to almost a GHz, where benchmarking performance can be expected for bandwidths on the order of hundreds of MHz. The requirements of our source design on the sample temperature and pump wavelength stability can be fulfilled with today's off-the-shelf components. 

We studied the purity of the photons produced with our source. Here, we made a distinction between the mode-excitation probability -- the probability that the dominant cavity resonance is excited -- and the spectral purity of the generated photons. We demonstrated that photons with overall purities exceeding 90\% can be generated when the design parameters and the pump spectral bandwidth are appropriately chosen for a wide range of photon bandwidths. 

Our source combines resonance enhancement as well as spectral redistribution of the PDC emission, which facilitates a more than 1000-fold increase in spectral source brightness at the memory transition when compared to a standard non-resonant PDC with external filtering. This is particularly appealing in light of large-scale quantum networks with many sources in parallel, since the overall energy consumption of such an architecture can be drastically reduced with our design. 

We also investigated the feasibility of fine-tuning the photon central frequency in our design. We found that tuning both sample temperature and pump frequency simultaneously facilitates a smooth and well-controlled tuning of the photon central frequency over a complete free-spectral range of the resonant structure. In combination with the coarse tuning provided by tuning pump or temperature only, our source design can be adapted exactly to a desired memory transition.

Finally, after having concentrated on our demonstrator source, we presented the adaptability of our design to different quantum memories. With slight changes in the source geometry, our design can address memory transitions of diverse material systems, for instance rare-earth ions or atomic vapours. We expect our source to become a part of many future hybrid quantum networks that seek to combine photons and solid state systems. 


\begin{acknowledgements}
This work was supported by the European Commission through the project QCUMbER (project reference: 665148).
\end{acknowledgements}

\bibliographystyle{spphys}       

\end{document}